\documentclass[12pt,prd,showpacs]{revtex4}
\usepackage{amssymb,amsfonts,amsmath}
\makeatletter
\renewcommand{\@makefntext}[1]{\parindent=1em\noindent\hbox to 1.8em{\hss$^{\@thefnmark}$}#1}
\renewcommand{\@footnotemark}{\hbox{\mathsurround=0pt$^{\@thefnmark}$}}
\newcommand{\ftnote}[2]{\footnotemark[#1]\footnotetext[#1]{#2}}
\makeatother

\begin{document}

\title{Chiral restoration in excited nucleons versus $SU(6)$}

\author{L. Ya. Glozman}
\affiliation{Institute for Physics, Theoretical Physics Branch, University of Graz, Universit\"atsplatz 5, A-8010 Graz, Austria}

\author{A. V. Nefediev}
\affiliation{Institute of Theoretical and Experimental Physics, 117218, B.Cheremushkinskaya 25, Moscow, Russia}

\begin{abstract}
We compare axial charges of excited nucleons,
as predicted by the chiral symmetry restoration picture, with
the traditional, moderately successful for the ground-state baryons $SU(6)$ symmetry. The axial charges of excited
nucleons can (and will) be measured in lattice QCD simulations, and comparison
of the lattice results with the two different symmetry schemes will
give an insight on the origins of the excited hadron masses as well as on
interrelations of chiral symmetry and confinement.
\end{abstract}

\maketitle

\section{Introduction}
There are well-known phenomenological successes of the $SU(6)$ flavour-spin symmetry in description
of static properties of the ground states of baryons \cite{GR,K}. Indeed, all the $1/2^+$ and $3/2^+$ ground
states in
the $u$-, $d$-, and $s$-quark sector, that is the octet and the decuplet of $SU(3)$, fall into the {\bf 56}-plet of $SU(6)$.
Splittings within this {\bf 56}-plet are due to the $SU(3)$ breaking, that is due to different masses of the
$u$, $d$, and $s$ quarks, as well as because of the $SU(6)$ breaking via the effective spin and/or flavour-spin interactions.
Magnetic moments (more exactly, their ratios) of the ground states typically agree,
within 10-15\%, with the $SU(6)$ predictions. The axial charge of the nucleon, as predicted by the $SU(6)$, $G^A_N = 5/3$, is
essentially larger than the empirical value, $G^A_N = 1.26$. Possible reasons for
the deviation are: relativistic effects, $SU(6)$ breaking, pion cloud effects, etc. Taking these effects into account allows one to improve the
$SU(6)$ prediction for the $G^A_N$.

It is established that indeed, in the large-$N_C$
limit of QCD, ground states satisfy the contracted $SU(6)$ algebra \cite{GS,DM}, and both the
constituent quark and the Skyrme descriptions become algebraically equivalent. While we do not know
yet all dynamical details, we understand that the emergency of the $SU(6)$ symmetry is a consequence
of the spontaneous breaking of chiral symmetry in QCD: at low momenta the valence quarks acquire
a rather large dynamical mass due to their coupling to the quark condensate of the QCD vacuum.
Whatever particular gluonic interaction in QCD is responsible for chiral symmetry breaking,
the microscopical mechanism of this breaking is related to the selfenergy dressing of quarks. We also know
that this selfenergy (``constituent quark mass") is not a constant, because at large space-like momenta
the asymptotic freedom of QCD requires this selfenergy to vanish. What is its fate at the time-like momenta
depends on microscopical details of confinement and chiral symmetry breaking and is unknown.

Successes of the $SU(6)$ symmetry for excited baryons are much more moderate. For the description of excited
states the $SU(6)$ must be supplemented by a dynamical assumption that specifies the radial and orbital structure of baryons.
In the simplest case of the harmonic oscillator confining potential in a nonrelativistic $3q$ system, the energy is totally fixed
by the principal quantum number $N$, which is the number of excitation quanta in the system.
Then the excited nucleon and delta states of negative parity in the $1.5\div 1.7$~GeV region
belong to the $N=1$ shell and have the flavour-spin structure prescribed by the
{\bf 70}-plet of the $SU(6)$. The amount of the empirical negative parity states in this energy region
as well as their quantum numbers nicely fit the $SU(6)\times O(3)$ classification, which is quite a nontrivial
prediction, indeed. Historically it was taken as a justification for the constituent quark model for excited states.

This scheme leads to a pronounced gap between the $N=0$ and $N=1$ shells, of the order of $500\div 700$~MeV.
One should expect then that plenty of the positive-parity
states which, by parity, could belong only to the $N=2$ shell, should be well above the $N=1$
shell and lie in the region of around 2~GeV. In reality, however, the positive-parity states have roughly
the same excitation energies as the negative-parity states. Even more, some of them (the Roper-like
states) lie below the negative-parity resonances.
The $SU(6)\times O(3)$ scheme predicts three different supermultiplets
in the $N=2$ shell: a {\bf 56}-plet, a {\bf 70}-plet, and a {\bf 20}-plet, which is a vast amount
of states. Only a handful of them are observed experimentally. Although it is possible to engineer
a $SU(6)$ breaking interaction that provides a lowering of some of the positive-parity states
from the $N=2$ shell \cite{GLR}, still the amount of the observed positive-parity states is much smaller than
prescribed by the $SU(6)\times O(3)$ symmetry.

Phenomenologically the positive- and negative-parity excited states almost systematically
form approximate parity doublets. From the $SU(6) \times O(3)$ point of view such a doubling is unnatural and
accidental. If it is not accidental, then a symmetry must be behind such a parity doubling. At the same time
there must be reasons for this symmetry not to be explicit in the lowest baryons. It was
suggested that this parity doubling reflects in fact a restoration in excited hadrons of the spontaneously
broken chiral symmetry of QCD (effective chiral and $U(1)_A$ restorations) \cite{G1,CG,G2,G3,GNJL1,WG,CG2,C}
--- see Ref.~\cite{G4} for a review.
The chiral symmetry restoration in excited nucleons requires that these hadrons must decouple from the Goldstone bosons and
that their axial charges must vanish \cite{G4,GN,JPS}. At the same time it forbids $N^* \rightarrow N \pi$ decays \cite{G5}.
Experimentally, the $N^*N\pi$ coupling constants are strongly suppressed for all observed approximate
parity doublets, indeed.  There is only one state, $3/2^-,N(1520)$, whose chiral partner is certainly missing in the
spectrum. This state does strongly decay into $N\pi$. Hence one observes a 100\% correlation of the decay data
with the spectroscopic parity doublets, as predicted by chiral restoration \cite{G5}. Although the diagonal axial charges
of excited states cannot be measured experimentally, they can be studied on the lattice. The first lattice results
for the lowest negative-parity state, $1/2^-,N(1535)$ (in this case a possible chiral partner is the Roper
resonance $1/2^+,N(1440)$) show a very small axial charge thus supporting chiral restoration \cite{TK}.

Chiral restoration in excited hadrons means that a mass generation mechanism in these hadrons is essentially
different compared to the lowest states. In the latter, the mass is driven by chiral symmetry breaking
in the vacuum, that is by the quark condensate. Consequently, the pion coupling to the valence quarks must also be
important for this mass generation. Chiral symmetry is strongly broken in these states and is
realised nonlinearly \cite{W,JPS}. In contrast, according to the effective chiral restoration scheme, the quark
condensate in the vacuum is almost irrelevant for the approximate parity doublets and their mass  has mostly a
chirally symmetric origin.

This issue of  mass generation and interrelation between chiral symmetry breaking
and confinement is a key for understanding the QCD phase diagram. If
chiral symmetry is indeed approximately restored in the excited hadrons, where  physics
should be dominated by confinement, then it is quite likely that, above the chiral
symmetry restoration point at finite chemical potential, one would have a new phase \cite{LP}
that represents a confining but chirally symmetric matter \cite{GW},
rather than a deconfining quark matter.
If so, there will be dramatic
implications for the QCD phase diagram and astrophysics.

Within the $SU(6) \times O(3)$ scheme and within related constituent quark models there are no
chiral partners at all. Consequently, the whole hadron mass within this picture has to be due to chiral symmetry breaking in
the vacuum. Indeed, the only scenario that allows the $u,d$-hadrons in the chiral limit not to have chiral partners requires
their masses to originate from the chiral symmetry breaking in the vacuum \cite{GL,NJL,LEE}.
Then the axial properties of hadrons must be essentially different as compared to the chiral
restoration picture. Consider, for example, the pion decay properties of the lowest $1/2^-$ state, $N(1535)$, in the framework of
the $SU(6) \times O(3)$ scheme and related constituent quark models. A very small
$N(1535)\to N \pi$ decay width can be achieved in this scheme if one assumes the $N(1535)$ wave function to be a superposition of
two $70$-plet states with different spin configurations, corresponding to the total quark spin
$S=1/2$ and $S=3/2$, which are mixed by the $SU(6)$-breaking spin--spin and spin--tensor quark--quark forces \cite{IsKarl,KI}.
Such a superposition provides a cancellation of two, individually large, contributions into the
$N(1535) \to N \pi$ decay width. At the same time, such a superposition of two configurations
induces an enhancement of the $N(1535) \to N \eta$ constant. What is important, however, is that the orthogonal
combination of these two $S=1/2$ and $S=3/2$ states represents then the wave function of the next excited
$1/2^-$ state, $N(1650)$. It has to come as a mystery then that the decay coupling constant for the transition $N(1650) \to N \pi$ is as small as that
for the decay $N(1535) \to N \pi$ \cite{G5}\ftnote{1}{Note that typically the constituent quark models \cite{CR}, as
well as the large-$N_C$ expansions \cite{JS}, operate with the nonrelativistic decay amplitudes and, as a result, with
improper phase--space factors. They try to fit, with some free parameters, decay widths, rather than coupling constants.
Physics is contained, however, in the coupling constants. Attempts to describe strong decays
within relativistic constituent quark models without fitting lead to the results that are qualitatively
incompatible with phenomenology \cite{M}.}. Furthermore, why are both these coupling constants much smaller than the $\pi NN$ one?
On the contrary, why are the decay constants for the would-be $LS$ partners (within the $SU(6) \times O(3)$) $1/2^-,N(1535)$ and $3/2^-,N(1520)$
so dramatically different? Why is there a clear correlation of the decay patterns with the approximate
parity doublets? We are not aware of any satisfactory explanation of these facts within the $SU(6) \times O(3)$ scheme.

The aim of the present paper is to perform a systematic comparison of the predictions of the $SU(6) \times O(3)$ and the chiral restoration scheme for
the axial charges.
While they cannot be measured experimentally, they are a subject of intensive lattice calculations.
A comparison of different symmetry predictions with the future lattice results will provide
a clue for our understanding of the mass generation mechanism in excited hadrons as well as
interrelations of chiral symmetry and confinement.

In the next section we review predictions of chiral symmetry restoration for
diagonal and off-diagonal axial charges. The third section is devoted to
the same constants evaluated within the $SU(6) \times O(3)$ symmetry scheme. In the discussion/conclusion
part we compare both predictions and give an outlook.

\section{Axial charges for chiral parity doublets}

In this section we review the results for the axial charges of chiral parity doublets
\cite{G4}. Assume that we have a free $I=1/2$ chiral doublet $B$ in the $(0,1/2)\oplus(1/2,0)$
representation and there are no chiral symmetry breaking terms. This doublet is a column \cite{LEE},
\begin{equation}
B=\genfrac{(}{)}{0pt}{0}{B_+}{B_-},
\label{doub}
\end{equation}
where the bispinors $B_+$ and $B_-$ have a positive and a negative parity, respectively.
The axial transformation law for the $(0,1/2)\oplus (1/2,0)$ representation
provides a mixing of the fields $B_\pm$\ftnote{2}{Note that
the axial transformation given in Ref.~\cite{LEE} is incorrect as it
breaks chiral symmetry of the kinetic term. The correct axial
transformation is given in Ref.~\cite{G4}.}:
\begin{equation}
B\to\exp\left(i\frac{\theta^a_A\tau^a}{2}\sigma_1\right)B.
\label{VAD}
\end{equation}
Here $\sigma_1$ is a Pauli matrix that acts in the  $2 \times 2$ space of the parity doublet.
Then the chiral-invariant Lagrangian of the free parity doublet is given as
\begin{equation}
\mathcal{L}_0 =i\bar{B}\gamma^\mu\partial_\mu B-m_0\bar{B}B
=i\bar{B}_+ \gamma^\mu \partial_\mu B_+ +i\bar{B}_-\gamma^\mu\partial_\mu B_-
-m_0\bar{B}_+B_+-m_0\bar{B}_-B_-.
\label{lag}
\end{equation}
Alternative forms of this Lagrangian can be found in Refs.~\cite{DTK,TIT}.

A crucial property of this Lagrangian is that the fermions
$B_+$ and $B_-$ are strictly degenerate and have a nonzero chiral-invariant mass $m_0$. In contrast, for
usual fermions, chiral symmetry in the Wigner--Weyl mode restricts particles to be massless. Thus there are two possibilities to
satisfy chiral symmetry with the Dirac-type fermions:
\begin{enumerate}
\item The standard scenario, which is to be considered for the nucleon and other ground-state baryons in the light-quark sector:
(i) fermions are massless in the Wigner--Weyl mode;
(ii) independent chiral partners are not required;
(iii) fermion mass can be generated in the Nambu--Goldstone mode only due to spontaneous breaking
of chiral symmetry in the vacuum, that is via the coupling of the fermion with the chiral
order parameter (quark condensate).
\item The chiral symmetry restoration scenario, which applies to highly excited hadrons:
(i) parity-doubled fermions are massive already in the Wigner--Weyl mode;
(ii) this mass is manifestly chiral-invariant and is not related at all to the quark condensate;
(iii) the role of the chiral symmetry breaking in the Nambu--Goldstone mode (that is of the quark condensate)
is to lift the chiral degeneracy of the opposite-parity baryons. Effective chiral
restoration means that these opposite-parity baryons almost entirely decouple from the
quark condensate and most of their mass is manifestly chiral-invariant;
(iv) consequently there appear approximate parity doublets in the spectrum.
\end{enumerate}

The global chiral symmetry properties  (\ref{VAD}) of the Lagrangian  (\ref{lag}) imply,
via the N{\"o}ther theorem, the following form of the conserved axial-vector current:
\begin{equation}
A^a_\mu=\bar{B}_+\gamma_\mu\frac{\tau^a}{2}B_-+\bar{B}_-\gamma_\mu\frac{\tau^a}{2}B_+.
\label{ac}
\end{equation}
This current does not contain diagonal terms, like $\bar{B}_+\gamma_\mu \gamma_5\frac{\tau^a}{2}B_+$ or
$\bar{B}_-\gamma_\mu \gamma_5 \frac{\tau^a}{2}B_-$.
Consequently, the diagonal axial charges of the parity-doubled baryons $B_+$ and $B_-$ are exactly 0. In contrast, the
off-diagonal axial charges, which normalise axial transitions between the members
of the parity doublet are exactly 1. Hence general chiral symmetry properties
of the chiral parity doublets uniquely fix the axial properties of the opposite-parity baryons:
\begin{equation}
G^A_{+}=G^A_{-}=0,\quad G^A_{+-}=G^A_{-+}=1.
\label{ach}
\end{equation}
This is another crucial property that distinguishes parity doublets from the usual Dirac fermions, the latter having $G^A = 1$.
Any microscopic model of chiral parity doublets must satisfy these general constraints
(see Ref.~\cite{NRS0} for a particular microscopic realisation of these conditions in the framework of the Generalised Nambu--Jona-Lasinio model).

Consider now constraints implied by the conservation of the axial-vector current in
the chiral limit. Since the diagonal axial charges of the $B_+$ and $B_-$ strictly vanish,
then the Goldberger--Treiman relation, $g_{\pi B_\pm B_\pm}=\frac{G^A_\pm m_\pm}{ f_\pi}$,
 requires that the diagonal coupling constants to the
pion must vanish, $g_{\pi B_+B_+} = g_{\pi B_-B_-} = 0$. Hence small
values of the diagonal axial charges as well as pion--baryon coupling constants taken together
with the large baryon mass would tell us that the origin of this mass is not
due to chiral symmetry breaking in the vacuum. It is a challenge for lattice calculations to measure these quantities for excited
states \cite{TK}.

A similar relation can be obtained for the off-diagonal coupling to the pion, $g_{\pi B_+B_-} = 0$. Indeed,
a generic axial-vector current matrix element between two arbitrary $1/2^+$ and $1/2^-$ isodoublet
baryons is given as:
\begin{equation}
\langle B_-(p_f)|A^a_\mu|B_+(p_i)\rangle=\bar{u}(p_f)\left[\gamma_\mu H_1(q^2)+
\sigma_{\mu \nu}q^\nu H_2(q^2)+q_\mu H_3(q^2)\right]\frac{\tau^a}{2}u(p_i),\quad q=p_f-p_i,
\label{fff}
\end{equation}
where $H_1$, $H_2$, and $H_3$ are formfactors. Then the matrix element of the divergence of
the axial-vector current is
\begin{equation}
\langle B_-(p_f)|\partial^\mu A^a_\mu|B_+(p_i)\rangle=i\left[(m_+-m_-)H_1(q^2)+q^2H_3(q^2)\right]\bar{u}(p_f)\frac{\tau^a}{2}u(p_i).
\label{df}
\end{equation}
Because of the axial-vector current conservation, this matrix element must vanish. There
are two possibilities to satisfy this:
\begin{enumerate}
\item $m_+ \neq m_-$. Then the Goldstone boson pole is required in $H_3(q^2)$ and one arrives at
a generalised Goldberger-Treiman relation:
\begin{equation}
g_{\pi B_+B_-}=\frac{G^A_{+-}(m_+ - m_-)}{2 f_\pi},\quad G^A_{+-}=H_1(0).
\label{ggt}
\end{equation}
\item $m_+ = m_-$. Then $H_3(q^2)=0$ and, automatically, $g_{\pi B_+B_-}=0$.
\end{enumerate}

Note that the condition $g_{\pi B_+B_-} = 0$ is a general consequence of the axial-vector current
conservation for degenerate opposite-parity baryons. Whatever reason for this degeneracy
is --- whether it is due to chiral restoration or something else --- the off-diagonal pion coupling must vanish.

In reality, of course, chiral symmetry is never completely restored in excited baryons.
The effect of chiral symmetry breaking in the vacuum is to split the masses of $B_+$ and $B_-$.
Consequently the axial charges should deviate from the limiting values given in Eq.~(\ref{ach}).
However, if the coupling to the condensate is weak, then chiral symmetry breaking is
only a small perturbation, and one should expect that, for approximate
parity doublets, the relations (\ref{ach}) are approximately satisfied.
A possible assignment of excited nucleons to the chiral multiplets is given in Table~\ref{t1}.

\begin{table}[t]
\caption{Chiral multiplets of excited nucleons.
Comments: (i) All these states are well established and
can be found in the Baryon Summary Table of the Review of Particle
Physics. (ii) There are two possibilities to assign the chiral representation:
$(1/2,0) \oplus (0,1/2)$ or $(1/2,1) \oplus (1,1/2)$ because
there is a possible chiral pair in the $\Delta$ spectrum
with the same spin with similar mass. (iii) The missing  chiral partner
is predicted.}
\begin{ruledtabular}
\begin{tabular}{llll}
Spin & Chiral multiplet &  Representation  & Comment\\
1/2& $N_+(1440 ) - N_-(1535)$ & $(1/2,0) \oplus (0,1/2)$  & (i) \\
1/2& $N_+(1710) - N_-(1650)$ & $(1/2,0) \oplus (0,1/2)$  & (i) \\
3/2& $N_+(1720) - N_-(1700)$ & $(1/2,0) \oplus (0,1/2)$  & (i) \\
5/2&$N_+(1680) - N_-(1675)$ & $(1/2,0) \oplus (0,1/2)$ & (i) \\
7/2&$N_+(?) - N_-(2190)$ &  see comment (ii)   & (i),(ii),(iii) \\
9/2&$N_+(2220) - N_-(2250)$ & see comment (ii) & (i),(ii) \\
11/2&$N_+(?) - N_-(2600)$ &   see comment (ii)  & (i),(ii),(iii) \\
\hline
3/2& $ N_-(1520)$ & no chiral partner  &  (i)
\end{tabular}
\end{ruledtabular}
\label{t1}
\end{table}

\section{Axial charges from the $SU(6)$ symmetry.}
First, let us review briefly the quantum numbers and the $SU(6)$ symmetric wave functions
for excited baryons, known from the first years of the quark model.  The flavour-spin
$SU(6)$ multiplet is uniquely specified by the corresponding Young pattern $[f]_{FS}$. In particular,
the ${\bf 56}$-plet is given by the completely symmetric Young pattern $[3]_{FS}$, the ${\bf 70}$-plet
is given by the mixed symmetry pattern $[21]_{FS}$, and the ${\bf 20}$-plet is specified by the
antisymmetric Young diagram, $[111]_{FS}$. Different baryons within a given $SU(6)$ multiplet
are characterised by the flavour $SU(3)_F$ and spin $SU(2)_S$ symmetries. In particular, the
decuplet, octet, and singlet states have the following Young patterns: $[3]_F$, $[21]_F$, and $[111]_F$,
respectively. The spin symmetry $[f]_S$ is uniquely determined by the total spin $S$ in the $3q$ system,
that is for spins $S=3/2$ and $S=1/2$ it is $[3]_S$ and $[21]_S$, respectively. The orbital state in
the $3q$ system is specified in general by the orbital angular momentum $L$ as well as by the spatial
permutational symmetry $[f]_X$. The latter is uniquely fixed by the Pauli principle, $[f]_X= [f]_{FS}$.
Within a specific spatial basis there can appear additional
quantum numbers in order to specify uniquely the given spatial basis function, as it happens, for example, to
the highly symmetric harmonic basis. In the latter case, these additional quantum numbers include
$N$, which specifies the energy and is the number of the
excitation quanta (it is the principal quantum number for the $U(6)$ spatial symmetry of the harmonic
oscillator in a three-body system), and the {\em spatial} $SU(3)$ symmetry of the orbital wave function
is fixed by the symbol $(\lambda \mu)$.
Given all these quantum numbers one traditionally assigns excited nucleons of positive and negative parity
to the $SU(6)$ multiplets, as is shown in Table~\ref{t2}.

\begin{table}[t]
\caption{The $SU(6)$ assignments of the well established  nucleons below 2~GeV. Vacant states from the {\bf 20}-plet
are not shown. }
\begin{ruledtabular}
\begin{tabular}{ll}
$N(\lambda\mu)L[f]_X[f]_{FS}[f]_F[f]_S$& LS multiplet \\
$0(00)0[3]_X[3]_{FS}[21]_F[21]_S$ & $\frac12^+, N$ \\
$2(20)0[3]_X[3]_{FS}[21]_F[21]_S$ & $\frac12^+, N(1440)$\\
$1(10)1[21]_X[21]_{FS}[21]_F[21]_S$ & $\frac12^-, N(1535);\frac32^-,N(1520)$\\
$1(10)1[21]_X[21]_{FS}[21]_F[3]_S$ & $\frac12^-, N(1650);\frac32^-,N(1700);\frac52^-,N(1675)$ \\
$2(20)2[3]_X[3]_{FS}[21]_F[21]_S$&$\frac32^+,N(1720);\frac52^+,N(1680)$\\
$2(20)0[21]_X[21]_{FS}[21]_F[21]_S$ & $\frac12^+, N(1710)$ \\
$2(20)0[21]_X[21]_{FS}[21]_F[3]_S$ & $\frac32^+, N(?)$\\
$2(20)2[21]_X[21]_{FS}[21]_F[21]_S$ & $\frac32^+, N(?);\frac52^+,N(?);$ \\
$2(20)2[21]_X[21]_{FS}[21]_F[3]_S$ & $\frac12^+, N(?);\frac32^+,N(?);\frac52^+,N(?);\frac72^+,N(?)$\\
\end{tabular}
\end{ruledtabular}
\label{t2}
\end{table}

Our purpose now is to calculate the diagonal and off-diagonal axial charges for
excited baryons that can be accessed on the lattice. The $SU(6)$ symmetry is a nonrelativistic symmetry. Then the diagonal and
off-diagonal axial charges for the $SU(6)$ baryons are given by the matrix elements of the nonrelativistic single-quark axial-vector
charge operators,
\begin{equation}
G^A_{fi}=\langle\Psi_f(1,2,3)|\sum_{n=1}^3Q^A_n|\Psi_i(1,2,3)\rangle.
\label{me}
\end{equation}
The nonrelativistic leading order axial charge operator $Q^A_n$ for a point-like Dirac constituent quark (index $n$ numerates quarks in the baryon)
is given by the Gamov-Teller operator $\sigma_3\tau_3$, where $\vec\sigma$ and $\vec \tau$ are the spin- and isospin operators, respectively.
Note that there is no dependence on the spatial coordinate in the 
nonrelativistic leading order axial charge
operator\ftnote{3}{Strictly speaking, the nonrelativistic single-quark axial transition operator does contain a spatial dependence, via
the exponent $\exp{[i\vec{q}\vec{r}]}$. In the meantime, the diagonal baryon axial charge is defined at the point $q^2=0$, that would
yield, in the nonrelativistic limit, $\vec{q}=0$.
In the case of the off-diagonal axial charge, the role played by the spatial dependence of the transition operator strongly depends on
the baryon masses in the initial and in the final states. Indeed, for the baryons with substantially different masses,
the off-diagonal transition is determined by the corresponding axial form-factor $H_1(q^2)$ --- see Eq.~(\ref{fff}) --- taken at the point
$q^2 \neq 0$. Notice that there is no unique and unambiguous form of the nonrelativistic axial charge operator in this case.
On the contrary, in the present work, we consider only the opposite-parity states which are approximately degenerate in mass.
Then the same point $\vec{q}=0$ is to be considered for the nonrelativistic axial charge operator.}.
Consequently only diagonal matrix elements with respect to the spatial quantum numbers can have nonzero
values. Therefore within the nonrelativistic $SU(6) \times O(3)$ symmetry all the off-diagonal axial charges for all
degenerate opposite parity states vanish identically, $G^A_{+-}=G^A_{-+}=0$.  This is the first
crucial difference between the manifest $SU(6)\times O(3)$ and chiral symmetry of excited baryons.

It is a legitimate question to consider  relativistic corrections to the pure static nonrelativistic
axial transition operator. It is well known from the nuclear physics applications, long before
the naive quark model, that the lowest-order correction, $\sim 1/M $, is given by the operator
\begin{equation}
\frac{1}{2M}\vec{\sigma}({\vec p}_i + {\vec p}_f)\tau^a e^{i \vec q \vec r},
\label{cor}
\end{equation}
where ${\vec p}_i$ and ${\vec p}_f$ are the initial and final quark momenta. Consequently, if the
pure nonrelativistic (static) transition is forbidden for some reason, the contribution of
the relativistic corrections may become important. For example, this kind of operator was taken into account
in a four-parameter fit for the baryon resonance decays in Ref.~\cite{KI}, and its strength was actually
considered as a free parameter. Within the usual nonrelativistic constituent quark models,
however, $v/c \sim 1$, or even larger. Hence there are no reasons to consider only the first correction
and to ignore all others.

Note that the same operator leads to a nonvanishing $g_{\pi B_+ B_-}$.
This is in an obvious conflict with the most general requirement for the opposite-parity degenerate
baryons, $g_{\pi B_+ B_-}=0$ (see the text below Eq.~(\ref{ggt})), even for $G^A_{+-}\neq 0$.
It is not clear how could it be possible to satisfy this requirement within the naive quark model,
as a matter of principle, once different relativistic corrections are taken into account to the
axial transition operator. This issue represents an additional conceptual problem for the
naive quark model of excited parity-doubled baryons.

In lattice simulations it is possible to separate the lowest negative and positive parity $J=1/2$
states $N(1535)$, $N(1650)$, $N(1440)$, and $N(1710)$ \cite{BU}.
Given explicit $SU(6)$ wave functions --- see, for example, Ref.~\cite{GLR} --- it is straightforward to calculate the required diagonal
axial charges:
\begin{eqnarray}
\frac{1}{2}^+, N(1440):&\quad&G^A=\frac53,\\
\frac{1}{2}^+, N(1710):&\quad&G^A=\frac13,\\
\frac{1}{2}^-, N(1535):&\quad&G^A=-\frac19,\\
\frac{1}{2}^-, N(1650):&\quad&G^A=\frac59.
\end{eqnarray}
It is also possible to access on the lattice the lowest $3/2^-, N(1520)$ state \cite{LHC}. Its diagonal axial charge matrix element reads:
\begin{equation}
\langle 3/2^-,N(1520);M,T|\sum_{n=1}^3Q^A_n|3/2^-,N(1520);M,T\rangle=
\frac{\sqrt{5}}{3} C_{\frac32 M 10}^{\frac32 M}C_{\frac12 T 10}^{\frac12 T},
\label{n1520}
\end{equation}
where $M$ and $T$ are its total spin and isospin projection, respectively.

\section{Discussion and outlook}

Now we are in a position to confront systematically the predictions of the chiral restoration and
$SU(6)$ schemes for excited nucleons.  As we have already mentioned above, while
chiral symmetry restoration predicts the off-diagonal axial charges within chiral
multiplets to be close to 1 ($G^A_{+-}=G^A_{-+}\sim 1$), the $SU(6) \times O(3)$ requires all these off-diagonal axial charges
to vanish, $(G^A_{+-}=G^A_{-+}=0)$. For the diagonal axial charges the difference is also substantial. Indeed,
if the $1/2^+,N(1440)-1/2^-,N(1535)$ pair is the lowest approximate parity doublet, then
the diagonal axial charge of the lowest positive-parity state (the Roper state) must be small, $\sim 0$. On the contrary, the $SU(6)$ symmetry
suggests for this Roper state a large axial charge, the latter coincides with the nucleon axial charge.
The axial charge of the $N(1535)$ is predicted in both cases to be small. For the next approximate $J=1/2$ doublet,
the $SU(6)\times O(3)$ predicts for the negative-parity state, $1/2^-,N(1650)$ a rather
large diagonal axial charge and not a small diagonal charge for $1/2^+, N(1710)$. The chiral symmetry
restoration requires both of them to have a small axial charge. The state $3/2^-, N(1520)$ is a $LS$-partner
of the $1/2^-, N(1535)$ state within the $SU(6) \times O(3)$ symmetry.  According
to the chiral symmetry restoration scenario, there is no chiral partner for $3/2^-, N(1520)$, and hence its mass
is due to chiral symmetry breaking in the vacuum. In this case there are no strict constraints for its axial charge coming from
chiral symmetry. Naively one would expect its diagonal axial charge to be of the order 1, because empirically nucleon has
a rather large axial charge, though in actuality it can take any arbitrary value.

We have systematically compared diagonal and off-diagonal axial properties of the states which are expected to be
approached soon within the lattice QCD calculations. There are first unquenched QCD results for the diagonal axial charge of the $N(1535)$
resonance \cite{TK}, that is very small. Given a combined set of empirical spectroscopic and decay data,
discussed in the introduction, it provides an additional support for the chiral restoration picture. Still, specifically this
small axial charge is also marginally compatible with the $SU(6)$ picture. It would be very interesting to measure
diagonal and off-diagonal axial charges of other excited states. Such a program is now under way.

\begin{acknowledgments}
L. Ya. G. acknowledges support of the Austrian Science Fund through grant P19168-N16.
Work of A. N. was supported by the Federal Agency for Atomic
Energy of Russian Federation, by the grants RFFI-05-02-04012-NNIOa, DFG-436 RUS
113/820/0-1(R), and PTDC/FIS/70843/2006-Fi\-si\-ca, as well as
by the Federal Programme of the Russian Ministry of
Industry, Science, and Technology No. 40.052.1.1.1112,
and by the non-profit ``Dynasty" foundation and ICFPM.
\end{acknowledgments}

\end{document}